\documentclass[manuscript,3p,times,twocolumn,sort&compress,10pt]{elsarticle}
\usepackage{lineno,hyperref,amssymb,amsmath}
\usepackage{graphicx,multicol}
\usepackage{subfig}
\usepackage{float}
\usepackage{array}
\newcolumntype{C}[1]{>{\centering\arraybackslash}p{#1}}

\begin{document}

\makeatletter
\def\ps@pprintTitle{%
  \let\@oddhead\@empty
  \let\@evenhead\@empty
  \def\@oddfoot{\reset@font\hfil\thepage\hfil}
  \let\@evenfoot\@oddfoot
}
\makeatother

\begin{frontmatter}

\title{Neutron focusing in time and magnification of the time lens}

\author[Institute1]{A.I. Frank\corref{mycorrespondingauthor}}
\cortext[mycorrespondingauthor]{Corresponding author}
\ead{frank@nf.jinr.ru}

\address[Institute1]{Joint Institute for Nuclear Research, Dubna, Russia}

\begin{abstract}
The relation between the duration of the pulse of the neutron flux generated by the neutron source and the time image of this pulse formed by the time lens is considered. It is shown that the ratio of these magnitudes called “time magnification” depends not only on the geometric parameters of the device, as previously assumed, but also on the neutron velocity before and after the lens.
\end{abstract}

\begin{keyword}
ultra-cold neutrons, UCN source, time focusing, time magnification
\end{keyword}

\end{frontmatter}

\section{Introduction}
Most of the neutron sources, either existing or being designed, operate in a pulsed mode. In this case, the pulsed neutron flux density in the moderator of the source can exceed the average flux density by several orders of magnitude. This also fully applies to the ultracold neutron (UCN) flux, the energy spectrum of which occupies the lowest part of the Maxwellian spectrum of neutrons in the moderator. Therefore, the pulsed UCN flux from a thin moderator can be very significant. Thus, the question is how to take advantage of this circumstance.

A possible solution to the problem was proposed by F.L. Shapiro~\cite{Shapiro76}. It consists in filling the trap with UCNs only during the pulse and effectively isolating it for the rest of the time. In the ideal case of the absence of losses, the UCN density in the trap will be close to the peak neutron density, which may exceed the time average by several orders of magnitude.

The implementation of this idea is hindered by the fact that in practice the trap is remote from the moderator due to the presence of biological shielding. In this connection, there is a need for a several meters long transport neutron guide feeding the trap. The placement of the isolation valve near the moderator – the UCN source, leads to the fact that the neutron guide becomes part of the trap. Due to the small transverse size of the neutron guide, the frequency of neutron collisions against its walls is rather high, which greatly reduces the UCN storage time in the trap-neutron guide system and, accordingly, reduces the gain factor. Placing the valve at the entrance to the trap that is several meters away from the source is useful only in the case of sources with a low repetition frequency~\cite{Anghel09, Saunders13, Lauss14}. For sources with a repetition rate of several hertz or more, such as IBR-2 reactors~\cite{Ananiev77,IMFrank18, Aksenov09}, the designed IBR-3~\cite{Lopatkin20}, or the European Spallation Source~\cite{Garoby18}, the spread of the UCN flight times will exceed the intervals between pulses and using a valve at the entrance to the trap will be senseless.

This difficulty can be overcome by using a device that acts as a time lens and forms a time image of the source directly near the trap. It is assumed that the pulse duration of the time image can be of the same order as the duration of the true pulse flow of the UCNs in the source.

This proposal was formulated in~\cite{AIFrank00}, where the principles of forming a time image of a point source were discussed. Note that in the literature, the term "time lens" is used in relation to devices with significantly different functions. If in~\cite{AIFrank00} we were talking about a lens forming a given time distribution of the neutron flux at a known point in space,~\cite{Summhammer86, Baumann05} also considered the possibility of compressing the velocity interval at the observation point, that is, additional monochromatization of the neutron beam.

\begin{figure}[H]
	\centering
        \includegraphics[width=1.0\linewidth]{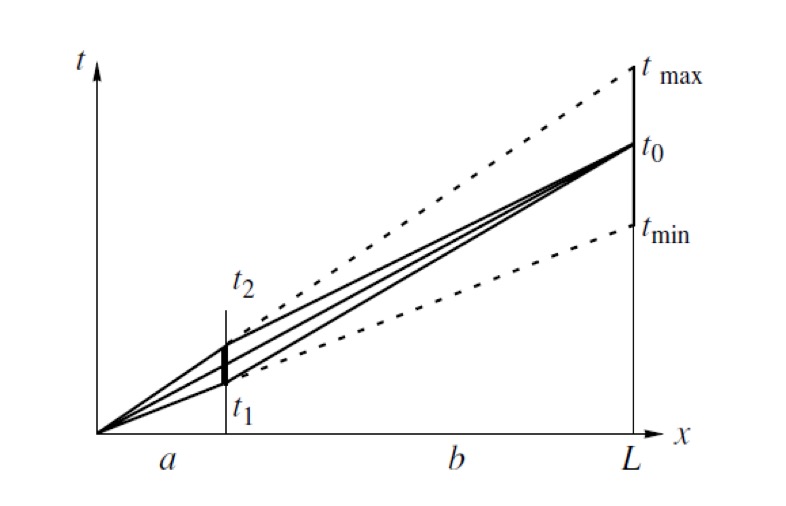}
 \caption
{
Time lens operation scheme.
}
\label{fig:fig1}
\end{figure}

Let us explain the idea of temporal focusing, following work~\cite{AIFrank00} (see Fig.~\ref{fig:fig1}).

 Let us assume that at the time instant $t=0$ neutrons are emitted from a point $x=0$ in the positive direction of the $X$-axis. The neutron velocities $v$ are distributed in a certain range of values. The time of their arrival $t_L$  at the observation point $x=L$ is distributed in the interval $t_1< t_L< t_2$. Supposedly, a time lens is located at the point  and it is capable of changing the neutron energy by an amount  according to a given time law in the interval . Therefore, the dependence  can be chosen in such a way that the velocities of neutrons, after passing through the lens, would satisfy the condition of simultaneous arrival at the observation point at the time instant $t_l=t_0$.

Assuming 
\begin{equation}\label{eq1}
\frac{a}{V_a} +\frac{a}{V_b},\;\;\; a+b=L,
\end{equation}
where $V_a$ and $V_b$ are neutron velocities before and after passing the lens, respectively, we immediately obtain
\begin{equation}\label{eq2}
\Delta E=\frac{m}{2} \left | \left(\frac{b}{t_0-t}\right)^2-\left(\frac{a}{t}\right)^2 \right |,\;\;\; t=\frac{a}{V_a},\;\;\; t_1<t<t_2,
\end{equation}
where $m$ is the neutron mass. 

Clearly, the pulse duration of a realistic source cannot be infinitesimal. Without discussing the issue of the pulse shape, we will now focus on the relation between the pulse duration of the source and its time image, that is, the issue of time magnification $M=\Delta/\tau$, where $\tau$ and $\Delta$ are the duration of the original pulse and its image, respectively. This issue is of paramount importance, since the ratio $\Theta=T/\Delta$ called “duty cycle” determines the relation between the values of the mean and pulsed UCN fluxes at the point $x=L$, where the conditional "consumer" of neutrons is located. Here $T$ is the repetition period of the source pulses.

Regarding the question of time magnification, in~\cite{AIFrank00} we limited ourselves to the statement that for a relatively small change in energy $\Delta E\ll E$, the formula of a thin lens is applied
\begin{equation}\label{eq3}
|M|=\frac{b}{a},
\end{equation}
without commenting on this result.

The question of the time magnification was also left unmentioned in the publications devoted to the results of the conducted experiments on the time focusing of UCNs~\cite{AIFrankPLA03, AIFrankJETPL03, Arimoto12, Imajo21}. At the same time, the issue becomes especially relevant in connection with the discussion of the possibility of creating a UCN source based on the principle of time focusing on the periodic action pulsed reactors in Dubna ~\cite{Aksenov09,Lopatkin20}. The purpose of this letter is to fill this gap.

\section{The duration of the pulse formed by the lens and the time magnification}
Let us assume that the source generates neutron flux pulses whose shape is determined by a certain function $f(t)$ with a maximum at $t=0$. In accordance with the above-stated, a time lens located at a point $x=a$ affects, one way or another, the neutrons reaching it, according to the time law in~\eqref{eq2}.

Let us recall that we are talking about a lens that, in accordance with~\eqref{eq2}, changes the energy of the neutrons trapped in it by an amount $E(t)$ at each moment of time $t$. Considering the case of a single pulse, we do not pay any attention to  the cyclic nature of the source operation at all. Taking this into account does not change anything in the subsequent calculations, except for the resulting time interval constraint $t_2-t_1\leq T$. Formula~\eqref{eq2} is obtained under the assumption that the neutrons were born at the moment of time $t=0$, that is, under the assumption that the velocity of the neutron in the first section of the flight is $V_a=a/t$. This calculated speed is denoted as $\tilde{V_a}$.

Relation ~\eqref{eq2}, which can be conveniently written as
 \setcounter{equation}{1}
\begin{subequations}
\begin{equation}
\left(V_a^2-V_b^2\right)=\left(\frac{a}{t}\right)^2 -\left(\frac{b}{t_0-t}\right)^2,\label{eq2:a}
\end{equation}
\end{subequations}
 \setcounter{equation}{3}
is valid for any neutrons that reach the lens at a time $t$, regardless of the velocity value $V_a$. In the ideal case of zero pulse duration $t=a/\tilde{V_a}$, $\tilde{V_b}=b/(t_0-t)$,  $t_0=(b/\tilde{V_b})+(a/\tilde{V_a})$, which is the condition of focusing. In the case of finite duration of the neutron pulse, $V_a$ may differ from $\tilde{V_a}$, if the neutron was not born exactly at the moment $t=0$.

Let us calculate the total flight time of a neutron born at an arbitrary moment of time $\delta$. If it reaches the lens at the moment of time $t$, the time of its flight along the path $a$ is $t_a=t-\delta$, and the velocity upon reaching the lens is
 \begin{equation}\label{eq5}
V_a=\tilde{V_a}\left(1+\frac{\delta}{t}\right),\;\;\;(\delta/t)\ll 1.
\end{equation}

From~\eqref{eq2:a} and~\eqref{eq5} it follows that the neutron velocity in section $b$ is
 \begin{equation}\label{eq6}
V_b=\tilde{V_b}\left(1+\frac{\delta}{t} \frac{\tilde{V_a}^2}{\tilde{V_b}^2}\right),
\end{equation}
and the time of flight of this section is
 \begin{equation}\label{eq7}
t_b=\frac{b}{\tilde{V_b}}\left(1-\frac{\delta}{t} \frac{\tilde{V_a}^2}{\tilde{V_b}^2}\right),\;\;\;\frac{\delta}{t} \frac{\tilde{V_a}^2}{\tilde{V_b}^2}\ll 1.
\end{equation}

There is an addition $\xi=\delta \left(b/a\right)\left(\tilde{V_a}/\tilde{V_b}\right)^3$ to the estimated time of flight $t_b=b/\tilde{V_b}$, and the ratio $\tilde{V_a}/\tilde{V_b}$ depends on the time $t$. In the symmetric case shown in Figure~\ref{fig:fig1}, the value $\left[1-\left(\tilde{V_a}^2/\tilde{V_b}^2\right)\right]$ changes the sign in the middle of the time interval $t_2-t_1$, when the neutron acceleration is replaced by deceleration.

\begin{figure}[H]
	\centering
        \includegraphics[width=1.0\linewidth]{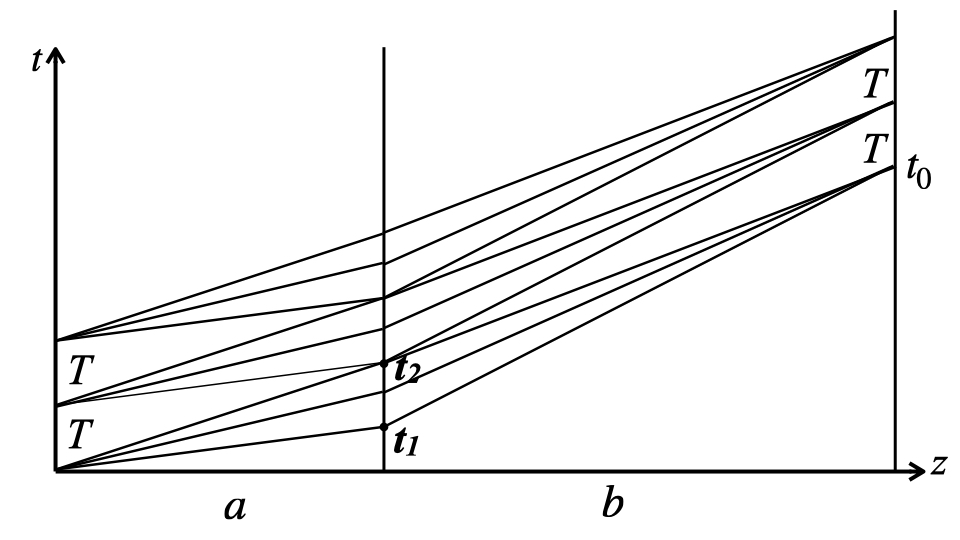}
 \caption
{
Operation scheme of a time lens that slows down neutrons.
}
\label{fig:fig2}
\end{figure}

Assuming that the velocity aperture is small, $t_2-t_1\ll t$ similar to the paraxial case in optics, remembering that in this case $t\approx t_a=a/V_a$, and averaging~\eqref{eq7} over the time interval $-\tau/2<\delta<\tau/2$, we will see that the pulse duration of the temporal image is $\Delta=\tau\left(b/a\right)$ and the magnitude of the time magnification is
\begin{equation}\label{eq8}
M=\frac{\Delta}{\tau}=\frac{b}{a},
\end{equation}
in accordance with~\cite{AIFrank00}. However, the case of the symmetric action of the lens is highlighted and it is quite possible to assume that the lens, possessing focusing properties, slows down or accelerates neutrons simultaneously (see Fig.~\ref{fig:fig2}).

Speculating in a similar way, and assuming, as above, the validity of the paraxial approximation, for the time increase we will obtain the following ratio
\begin{equation}\label{eq9}
M=\frac{\Delta}{\tau}=\frac{b}{a}\left(\frac{\tilde{V_a}}{\tilde{V_b}}\right)^3,
\end{equation}
where velocities $\tilde{V_a}$ and $\tilde{V_b}$ refer to some average "ray". 

Obviously, in the case where the time aperture is not too small, the lens not only transforms the pulse duration, but also distorts its shape. Thus, we come to the problem of aberrations. The shape of the resulting pulse can be calculated based on the known velocity distribution, the ratio $\tilde{V_a}/\tilde{V_b}$ for a particular time trajectory or "ray", and formula~\eqref{eq7}.

\section{Conclusion}
The idea of time focusing was formulated in~\cite{AIFrank00} as a possible way to restore a short pulse of the UCN flux generated by the source directly near the trap accumulating neutrons. Without focusing, the initial pulse is rapidly blurred during its transportation from the source to the trap due to velocity dispersion. At the same time, duration of the image formed by the time lens is of primary importance, since it is the ratio between the duration of the original pulse and its temporary image that determines the quality of this device.

Meanwhile, the question of the temporal magnification, that is, the relation between the duration of the original pulse and its image, remains practically unexplored. In ~\cite{AIFrank00}, relation~\eqref{eq9} given for the magnitude of the time increase is valid in the paraxial approximation. At the same time, it was overlooked that the case of symmetric focusing under discussion, when the average neutron velocity remains unchanged, is distinguished. In this paper, it is shown that in the general case of acceleration or deceleration of neutrons by a lens, the time magnification is proportional to the cube of the ratio of the final and initial velocities. For pulse compression, it is necessary to implement the geometry of a neutron accelerator. On the contrary, an attempt to combine time focusing with neutron moderation is inevitably associated with an increase in the pulse-image duration, which significantly devalues the very idea of temporal focusing.
\section*{Acknowledgments}
The author expresses his gratitude to M.A. Zakharov, G.V. Kulin and N.V. Rebrova for fruitful discussions.

\bibliographystyle{elsarticle-num} 

\bibliography{Frank_arxiv.bib}
\end{document}